\documentclass[12pt,a4paper]{article}
\pdfoutput=1

\usepackage[bulletsep]{collref}
\usepackage{amsmath,amssymb,graphicx}
\usepackage{pst-all}
\usepackage{bbm}

\makeatletter \@addtoreset{equation}{section} \makeatother

\makeatletter
\let\old@startsection=\@startsection
\let\oldl@section=\l@section
\renewcommand{\@startsection}[6]{\old@startsection{#1}{#2}{#3}{#4}{#5}{#6\mathversion{bold}}}
\renewcommand{\l@section}[2]{\oldl@section{\mathversion{bold}#1}{#2}}
\makeatother

\makeatletter
\let\old@makecaption=\@makecaption
\def\@makecaption{\small\old@makecaption}
\makeatother

\begin{document}

\thispagestyle{empty}
\begin{flushright}\footnotesize
\texttt{ITEP-TH-29/10}\\
\texttt{UUITP-25/10}\\ \vspace{0.8cm}
\end{flushright}

\renewcommand{\thefootnote}{\fnsymbol{footnote}}
\setcounter{footnote}{0}

\begin{center}
{\Large\textbf{\mathversion{bold} Holographic three-point functions \\ of semiclassical states
}
\par}

\vspace{1.5cm}

\textrm{K.~Zarembo\footnote{Also at ITEP, Moscow, Russia}}
\vspace{8mm}

\textit{CNRS -- Laboratoire de Physique Th\'eorique,
\'Ecole Normale Sup\'erieure\\
24 rue Lhomond, 75231 Paris, France }\\
\texttt{Konstantin.Zarembo@lpt.ens.fr}\vspace{3mm}

\textit{Department of Physics and Astronomy, Uppsala University\\
SE-751 08 Uppsala, Sweden}\\
\vspace{3mm}


\par\vspace{1cm}

\textbf{Abstract} \vspace{5mm}

\begin{minipage}{14cm}
We calculate the holographic  three-point functions in
$\mathcal{N}=4$ super-Yang-Mills theory in the case when two of the
operators are semiclassical and one is dual to a supergravity mode.
We further discuss the transition to the regime when all three
operators are semiclassical.
\end{minipage}

\end{center}

\vspace{0.5cm}


\newpage

\setcounter{page}{1}
\renewcommand{\thefootnote}{\arabic{footnote}}
\setcounter{footnote}{0}

\section{Introduction}

The correlation functions in  the $\mathcal{N}=4$  super-Yang-Mills
theory (SYM) can be calculated both at weak and at strong coupling,
owing to the AdS/CFT
duality\cite{Maldacena:1998re,Gubser:1998bc,Witten:1998qj}.  The
two-point functions, by conformal symmetry, are determined by the
spectrum of anomalous dimensions, which can be computed
non-perturbatively with the help of integrability (see
\cite{Rej:2009je,Serban:2010sr,Puletti:2010ge} for recent reviews),
in the leading planar order of the large-$N$ expansion. The
three-point functions are the simplest observables of the next order
in $1/N$. The three-point functions of protected operators can be
calculated at strong coupling in the supergravity approximation
\cite{Freedman:1998tz,Chalmers:1998xr,Lee:1998bxa,Arutyunov:1999en,Lee:1999pj}.
Non-protected operators with large quantum numbers constitute
another potentially solvable case.  At strong coupling they are dual
to classical spinning strings
\cite{Gubser:2002tv,Frolov:2003qc,Tseytlin:2003ii,Plefka:2005bk},
and there has been a renewed interest in computing their correlation
functions holographically
\cite{Buchbinder:2010gg,Janik:2010gc,Buchbinder:2010vw}, but going
beyond two-point correlators seems to be a difficult task. Here we
consider an intermediate case when two operators in the correlator
are semiclassical and one is protected\footnote{Similar calculations
are done, using a different method, in a parallel independent
publication \cite{Costa:2010rz}.}. The three-point function can then
be calculated with the help of the method developed in
\cite{Berenstein:1998ij}. We will also discuss transition to the
regime when all three states are semiclassical.

The three-point functions are determined by conformal symmetry up to
an overall numerical factor. If the operators are scalar and
conformal primary, their three-point functions have the form:
\begin{equation}\label{3pgen}
 \left\langle \mathcal{O}_1(x_1)\mathcal{O}_2(x_2)\mathcal{O}_3(x_3)\right\rangle
 =\frac{C_{123}}{|x_1-x_2|^{\Delta _1+\Delta _2-\Delta _3}
 |x_1-x_3|^{\Delta _1+\Delta _3-\Delta _2}
 |x_2-x_3|^{\Delta _2+\Delta _3-\Delta _1}
 }\,.
\end{equation}
The normalization here is important, and we will always assume  that the two-point functions of conformal primaries are unit normalized:
\begin{equation}\label{}
 \left\langle \mathcal{O}^\dagger _I(x)\mathcal{O}_J(y)\right\rangle=\frac{\delta _J^I}{|x-y|^{2\Delta _I}}\,.
\end{equation}
The constant $C^I_{JK}$ then determines the coefficient with which the operator $\mathcal{O}_I$ appears in the operator product of $\mathcal{O}_J$ and $\mathcal{O}_K$:
\begin{equation}\label{3OPE}
 \mathcal{O}_J(x)\mathcal{O}_K(0)=\sum_{I}^{}C^I_{JK}|x|^{\Delta _I-\Delta _J-\Delta _K}\mathcal{O}_I(0)+{\rm descendants}.
\end{equation}

\begin{figure}[t]
\centerline{\includegraphics[scale=0.6]{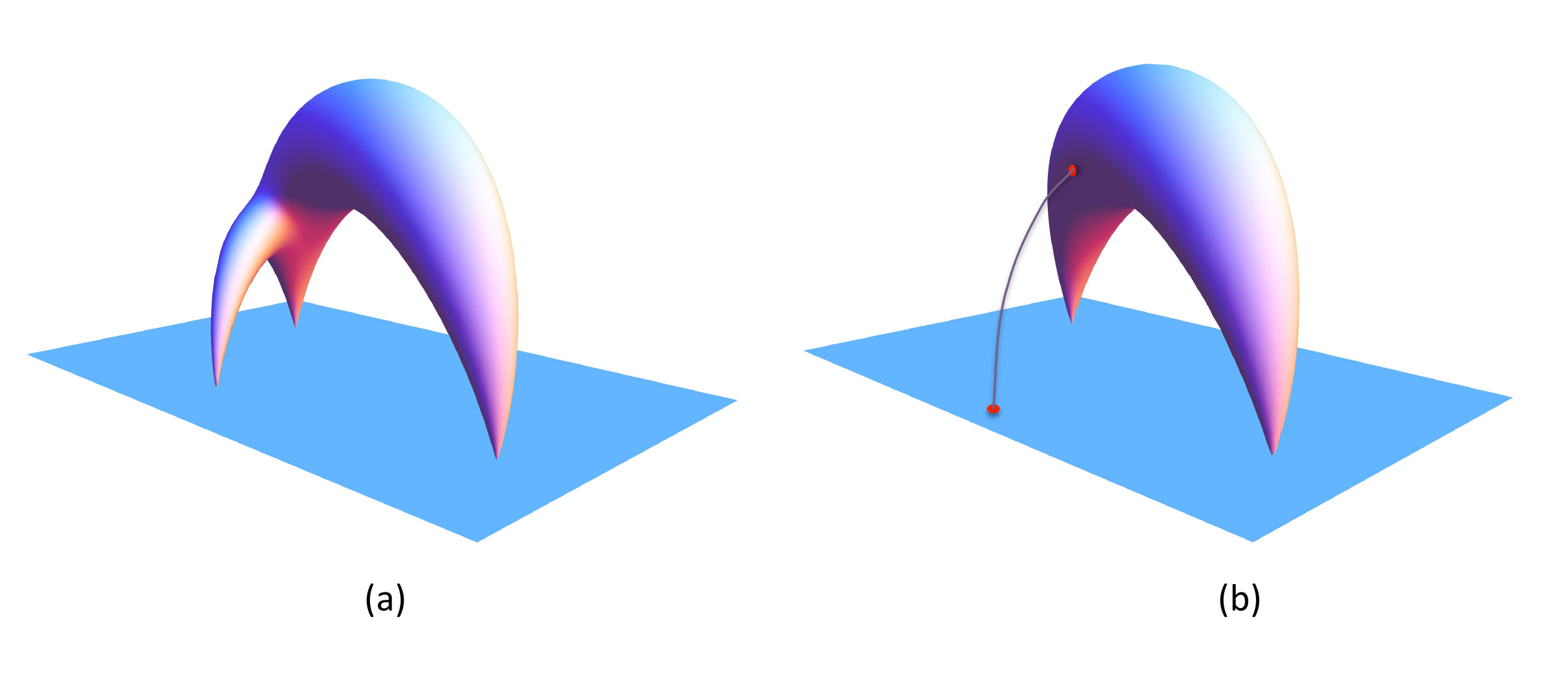}}
\caption{\label{StringDiagrms}\small (a) The three-point function of
semiclassical operators. (b) The correlator of two semiclassical
operators and a supergravity mode.}
\end{figure}

On the string side of the duality, a SYM correlator is calculated by
inserting vertex operators in the string path integral. The string
tension  $\sqrt{\lambda }/2\pi  $ is large when the SYM 't~Hooft
coupling $\lambda $ is large, and at strong coupling the path
integral is dominated by a saddle point. The semiclassical states
are characterized by parametrically large scaling dimensions,
$\Delta \propto\sqrt{\lambda }$, such that their vertex operators
should be taken into account as sources in the equations of motion
for the embedding coordinates in $AdS_5\times S^5$
\cite{PolyakovStrings2002}. The vertex operator insertions make the
string shrink and approach the boundary of $AdS_5$ at  $x_1$, $x_2$,
$x_3$, as shown in fig.~\ref{StringDiagrms}a\footnote{The
semiclassical picture of the holographic correlation functions has
been discussed at length in the context of the $AdS_3/CFT_2$
correspondence \cite{deBoer1998}.}.
The correlator  is determined by the classical string action, and is
thus exponential in $\sqrt{\lambda }$. The $x$-dependent factors in
(\ref{3pgen}) are consistent with this exponential dependence on
(since $\Delta _I\sim\sqrt{\lambda }$)  and, as shown in
\cite{Janik:2010gc}, the correct space-time structure of the two-
and three-point correlation functions indeed follows from the
classical string calculation. The OPE coefficients of the
semiclassical states should also be exponentially small (or large)
at strong coupling: $-\ln C^I_{JK}\propto \sqrt{\lambda }$.

The classical string solutions that are  dual to the two-point
functions are relatively simple. They can be obtained by the
Euclidean continuation of spinning strings in global AdS. In the
Poincar\'e patch the string worldsheet indeed collapses onto the
boundary at two points
\cite{Tsuji:2006zn,Buchbinder:2010gg,Janik:2010gc}. Constructing
solutions dual to three-point functions seems to be a difficult
problem\footnote{The classical decay process of certain folded
string solution has been studied in the Minkowski signature
\cite{Peeters:2005pb}. The relationship of these Minkowski-signature
solutions in global AdS to the three-point functions in SYM is not
clear to us. On the one hand,  as argued in \cite{Dobashi:2004nm},
the holographic correlation functions are described by tunneling.
The Euclidean signature in this respect is more natural
\cite{Buchbinder:2010vw}. On the other hand, the spinning string
solutions with non-zero angular momenta in AdS approach the boundary
at a point only if the worldsheet is Minkowskian
\cite{Janik:2010gc}, otherwise (in the Euclidean signature) the
boundary maps to a line \cite{Alday:2010ku,Buchbinder:2010vw}.}.
Here we study an intermediate case, when  two strings are "fat"
(dual to operators with $\Delta \propto\sqrt{\lambda }$) and one
string is "slim" and particle-like (dual to an operator with $\Delta
=O(1)$). The corresponding Witten diagram is shown in
fig.~\ref{StringDiagrms}b. We further assume that the "fat" string
is not much disturbed by the insertion of the "slim" vertex
operator, which means that two of the operators in the correlator
must be virtually the same\footnote{More precisely, they should be
conjugate.}. Then we only need to know the fat-string solution for
the two-point function and the vertex operator of the slim string.

\section{Holographic OPE}

One of the operators in the correlation functions  that we are going
to compute will always be a chiral primary. The chiral primary
operators in $\mathcal{N}=4$ SYM theory are symmetrized single-trace
products of the six scalar fields from the $\mathcal{N}=4$
multiplet:
\begin{equation}\label{}
 \mathcal{O}^{\rm CPO}_I=\frac{1}{\sqrt{k}}\left(\frac{8\pi ^2}{\lambda }\right)^{\frac{k}{2}}
 K^{i_1\ldots i_k}_I\mathop{\mathrm{tr}}\Phi _{i_1}\ldots \Phi _{i_k}.
\end{equation}
The coefficients $K^{i_1\ldots i_k}_I$ are symmetric  traceless
tensors of $SO(6)$, which at the same time define the spherical
functions on $S^5$:
\begin{equation}\label{spherical}
 Y_I(\mathbf{n})=K^{i_1\ldots i_k}_In_{i_1}\ldots n_{i_k}.
\end{equation}
We assume the following normalization condition:
\begin{equation}\label{}
 K^{i_1\ldots i_k}_IK^{i_1\ldots i_k}_J=\delta _{IJ}.
\end{equation}
It  guarantees the correct normalization of the two-point functions.
Requiring orthonormality of the spherical functions leads to a
different normalization condition, which is necessary to keep in
mind in the holographic calculations. The rescaling factor is
computed in \cite{Lee:1998bxa}.


\subsection{General formalism}

To compute the three-point functions of the chiral  primary
operators with semiclassical states, we will use the method applied
in  \cite{Berenstein:1998ij} to  correlation functions of a local
operator with a Wilson loop. The method can be generalized to any
non-local operator $\mathcal{W}$ which is dual to a classical string
worldsheet. This can be a Wilson loop as in
\cite{Berenstein:1998ij}, or a product of local operators as in our
case. Let us define the following ratio of correlation functions:
\begin{equation}\label{oW}
 \left\langle \mathcal{O}_I(x)\right\rangle_{\mathcal{W}}
 =\frac{\left\langle \mathcal{W}\,\mathcal{O}_I(x)\right\rangle}
 {\left\langle \mathcal{W}\right\rangle}\,,
\end{equation}
where  $\mathcal{O}_I$ is the local operator of interest, which is dual to a supergravity mode.

At large distances the correlator falls off as $|x|^{-2\Delta _I}$,
and we can extract the OPE coefficient
\begin{equation}\label{}
 \mathbb{C}_I[\mathcal{W}]=\lim_{x\rightarrow \infty }
 |x|^{2\Delta _I}\left\langle
 \mathcal{O}_I(x)\right\rangle_{\mathcal{W}}.
\end{equation}
Assuming that the operator $\mathcal{W}$ has a finite space-time support,
the OPE coefficient $\mathbb{C}_I[\mathcal{W}]$ determines the amplitude to find $\mathcal{O}_I$ in the operator
expansion of $\mathcal{W}$ around the origin:
\begin{equation}\label{}
 \mathcal{W}=\sum_{I}^{}\mathbb{C}_I[\mathcal{W}]\mathcal{O}_I(0)+{\rm
 descendants}.
\end{equation}
When $\mathcal{W}$ is a product of two  local operators, $\mathbb{C}_I$ is related to the conventional OPE coefficient from (\ref{3pgen}), (\ref{3OPE}):
\begin{equation}\label{andope}
 \mathbb{C}_I[\mathcal{O}^\dagger _J(x_1)\mathcal{O}_K(x_2)]=|x_1-x_2|^{\Delta_I }C^J_{IK}.
\end{equation}

How to calculate the correlation function (\ref{oW})
holographically? Keeping  in mind that  $\mathcal{O}_I$ is dual to a
supergravity mode and $\mathcal{W}$ is dual to a classical string,
the best suited formalism is a hybrid of the first-quantized string
theory and supergravity:
\begin{equation}\label{}
 \left\langle \mathcal{O}_I(y)\right\rangle_{\mathcal{W}}
 =\lim_{\varepsilon \rightarrow 0}\frac{\pi }{\varepsilon ^{\Delta _I}}\,
 \sqrt{\frac{2}{\Delta _I-1}}
 \left\langle  \phi_I (y,\varepsilon ) \,\frac{1}{Z_{\rm str}}\int_{}^{}\mathcal{D}\mathbb{X}\,\, {\rm e}\,^{-S_{\rm str}[\mathbb{X}]}\right\rangle_{\rm bulk}.
\end{equation}
The integration variables $\mathbb{X}$ in the  string path integral
are the embedding coordinates of the string worldsheet in
$AdS_5\times S^5$ and, in principle, fermions, but in the
semiclassical approximation the fermions will not be important. For
the AdS metric we take
\begin{equation}\label{}
 ds^2=\frac{dx_\mu ^2+dz^2}{z^2}\,,
\end{equation}
and parameterize $S^5$ by a unit 6d vector $\mathbf{n}$. The
boundary conditions on the string worldsheet are determined  by the
non-local operator $\mathcal{W}$. The supergravity field $\phi
_I(x,z)$ is dual  to the local operator $\mathcal{O}_I$, and the
bulk average is defined by the action (interactions in the bulk are
 $1/N$ suppressed):
\begin{equation}\label{}
 S_{\rm bulk} =\frac{1}{2}\int_{}^{}d^4x\,\frac{dz}{z^5}\,
 \left[\left(\partial \phi _I\right)^2+\Delta _I\left(\Delta _I-4\right)\phi
 _I^2\right].
\end{equation}
The field $\phi _I$ has the standard AdS propagator
\begin{equation}\label{bulkprop}
 \left\langle \phi _I(x,z)\,\phi _I(y,w)\right\rangle_{\rm bulk}
 =\frac{\Delta _I-1}{2\pi ^2}\,\,
 \frac{z^{\Delta _I}w^{\Delta _I}}{\left[z^2+\left(x-y\right)^2\right]^{\Delta
 _I}}\left(1+O(w^2)\right).
\end{equation}

The string action,
\begin{equation}\label{Sstr}
 S_{\rm str}=\frac{\sqrt{\lambda }}{4\pi }
 \int_{}^{}d^2\sigma \,
 \sqrt{h}h^{\mathbf{a}\mathbf{b}}\partial _\mathbf{a}\mathbb{X}^M
 \partial _\mathbf{b}\mathbb{X}^NG_{MN}+\ldots ,
\end{equation}
depends on the supergravity modes $\phi _I$ indirectly, through the disturbance of the metric created by the local operator insertion, which we denote by
$\gamma _{MN}$:
\begin{equation}\label{GMN}
 G_{MN}=g_{MN}+\gamma _{MN},
\end{equation}
where $g_{MN}$ is the unperturbed metric of $AdS_5\times S^5$. For
brevity we have omitted the coupling to $B_{MN}$, the dilaton, and
the fermions. The response of the metric to a perturbation in $\phi
_I$  depends on the details of the Kaluza-Klein reduction of the 10d
supergravity  on  $S^5$  \cite{Kim:1985ez}. In general, the 10d
metric perturbation is a linear combination of the normal modes
$\phi _I$ and their derivatives:
\begin{equation}\label{gammaMN}
 \gamma _{MN}=V^I_{MN}\phi _I,
\end{equation}
where $V^I_{MN}$ is a second-order differential operator with
$\mathbb{X}$-dependent coefficients.

At large $\lambda $ the string path integral is dominated by a saddle point, and we can substitute the classical solution  for $\mathbb{X}^M=(z(\sigma ),x^\mu (\sigma
),\mathbf{n}(\sigma ))$, expand the string action to the linear order in $\phi _I$, and use the propagator (\ref{bulkprop}) to calculate the bulk expectation value:
\begin{eqnarray}\label{}
 \left\langle \mathcal{O}_I(y)\right\rangle_{\mathcal{W}}
 &=&-\frac{\sqrt{2\left(\Delta _I-1\right)\lambda }}{8\pi ^2}
 \int_{}^{}d^2\sigma \,\sqrt{h}h^{\mathbf{a}\mathbf{b}}
 \partial _{\mathbf{a}}\mathbb{X}^M\partial _{\mathbf{b}}\mathbb{X}^N
 \nonumber \\
&& \times
 V_{MN}^I\left(\mathbb{X},\frac{\partial }{\partial x},\frac{\partial }{\partial z}\right)
 \frac{z^{\Delta _I}}{\left[z^2+\left(x-y\right)^2\right]^{\Delta _I}}\,,
\end{eqnarray}
The answer has the form of a vertex
operator in the coordinate representation
\cite{Polyakov:2001af,Tseytlin:2003ac} integrated over the classical string
worldsheet.

We can now take $y$ to infinity, to determine the OPE coefficient $
\mathbb{C}_I[\mathcal{W}]$. The vertex operator then simplifies a
bit.  In particular, $\partial /\partial x$ can be set to zero
\cite{Berenstein:1998ij}, because differentiating in $x$ increases
the power of $y$ and thus only contributes to correlators with the
descendants of $\mathcal{O}_I(y)$:
\begin{equation}\label{polufinal}
 \mathbb{C}_I[\mathcal{W}]=-\frac{\sqrt{2\left(\Delta _I-1\right)\lambda }}{8\pi ^2}\int_{}^{}d^2\sigma \,\sqrt{h}h^{\mathbf{a}\mathbf{b}}
 \partial _{\mathbf{a}}\mathbb{X}^M\partial _{\mathbf{b}}\mathbb{X}^N
 V_{MN}^I\left(\mathbb{X},\frac{\partial }{\partial z}\right){z^{\Delta _I}}.
\end{equation}
 Throughout this paper we will use  the conformal gauge for the classical solutions,
 and later the Weyl-invariant factor $\sqrt{h}h^{\mathbf{a}\mathbf{b}}$ will be replaced by $\delta ^{\mathbf{a}\mathbf{b}}$.

In the case when $\mathcal{O}_I$ is the chiral primary  operator
$\mathcal{O}_I^{{\rm CPO}}$, the dual supergravity field $s_I$ is a
mixture of the metric with the RR four-form. The metric perturbation
(\ref{gammaMN}) can be written explicitly after decomposing the 10d
index into the tangent-space indices of $AdS_5$ ($m,n,\ldots $) and
$S^5$ ($\alpha ,\beta ,\ldots $): $M=(m,\alpha )$. The Kaluza-Klein
reduction on $S^5$ \cite{Kim:1985ez} gives \cite{Lee:1998bxa} (see
also \cite{Arutyunov:1999fb}):
\begin{eqnarray}\label{fluctuatsii}
 h_{mn}&=&\frac{1}{\mathcal{N}_k}\,\,\frac{2}{k+1}\,Y_I\left[
 2\nabla_m\nabla_n-k(k-1)g_{mn}\right]s_I
\nonumber \\
h_{\alpha \beta }&=&\frac{2}{\mathcal{N}_k}\,kg_{\alpha \beta }Y_Is_I
\nonumber \\
a_{\alpha \beta \gamma \delta}&=&-\frac{1}{\mathcal{N}_k}\,\varepsilon _{\alpha \beta \gamma \delta \varepsilon }\nabla^\varepsilon Y_Is_I
\nonumber \\
a_{mnpr}&=&\frac{1}{\mathcal{N}_k}\varepsilon _{mnprs}Y_I\nabla^ss_I,
\end{eqnarray}
where $Y_I\equiv Y_I(\mathbf{n})$ are the spherical functions
(\ref{spherical}). The common normalization factor,
\begin{equation}\label{}
 \mathcal{N}^2_k=\frac{N^2k(k-1)}{2^{k-3}\pi ^2(k+1)^2}\,,
\end{equation}
takes into account the 10d gravitational constant,  which appears in
front of the supergravity action and is  equal to $\kappa
_{10}^2=(2\pi )^5/8N^2$ in the units of the AdS radius, the mixing
effects in the KK reduction, and the unusual normalization of the
spherical functions. The OPE coefficients owe their universal $1/N$
dependence on $N$ precisely to this normalization factor (and
 ultimately to the  10d gravitational constant).

 The RR field couples only to fermions which at the classical level can be
 neglected. Substituting the metric from (\ref{fluctuatsii}) into
 the string action
we finally get for  the OPE coefficient\footnote{In passing from
(\ref{polufinal}) to this equation the partial derivatives in $x$
can be dropped and the covariant derivatives $\nabla_\mu $ are
replaced by the Christoffel symbols.}
\begin{equation}\label{opeofcpo}
 \mathbb{C}_I^{\rm CPO}[\mathcal{W}]=
 \frac{2^{\frac{k}{2}-3}(k+1)\sqrt{k\lambda }}{\pi N}\int_{}^{}d^2\sigma \,Y_I(\mathbf{n} )
 \left[
 z^{k-2}\left(\partial x\right)^2-z^{k-2}\left(\partial z\right)^2-z^k
 \left(\partial \mathbf{n} \right)^2\right].
\end{equation}
We are going to use this formula to compute the three-point functions of various semiclassical operators with the BMN-type chiral primary,
\begin{equation}\label{trZ^L}
 \mathcal{O}_k=\frac{1}{\sqrt{k}}\left(\frac{4\pi ^2}{\lambda }\right)^{\frac{k}{2}}
 \mathop{\mathrm{tr}}Z^k,\qquad Z=\Phi _1+i\Phi _2.
\end{equation}
This operator is the highest-weight state in the $[0,k,0]$
representation of $SO(6)$. In the standard angular parameterization of the five-sphere,
\begin{eqnarray}\label{S5}
 \mathbf{n}&=&(\sin\theta \,\cos\varphi ,
 \sin\theta\,\sin\varphi,
 \cos\theta\,\sin\alpha\, \cos\psi,
 \cos\theta\,\sin\alpha\, \sin\psi,\nonumber \\ &&
 \cos\theta \,\cos\alpha\,\cos\beta,
 \cos\theta \,\cos\alpha \,\sin\beta),
\end{eqnarray}
the associated spherical function is
\begin{equation}\label{hw}
 Y_{k}=\left(\frac{n_1+in_2}{\sqrt{2}}\right)^k=2^{-\frac{k}{2}}
 \left(\sin\theta \right)^k\, {\rm e}\,^{ik\varphi  }.
\end{equation}

\subsection{BMN operators}

As a check on the formalism we first consider the case when the other two operators are also chiral primaries, but carry large quantum numbers. The OPE coefficients of three chiral primary operators are actually known exactly \cite{Lee:1998bxa}, since they are not renormalized and do not depend on $\lambda $ \cite{Lee:1998bxa,DHoker:1998tz,GonzalezRey:1999ih,Intriligator:1999ff,Eden:1999gh,Petkou:1999fv} (but there are non-trivial $1/N$ corrections \cite{Kristjansen:2002bb}). For the BMN primaries (\ref{trZ^L}) the OPE coefficients are equal to
\begin{equation}\label{}
 C^{k_1}_{k_2k_3}=\frac{1}{N}\,\sqrt{k_1k_2k_3}\,
\end{equation}
where the condition $k_1=k_2+k_3$ must be imposed to satisfy the R-charge conservation.

We can compute $ C^{J+k}_{Jk}$ at strong coupling in the regime when two operators are fat and one is slim:
\begin{equation}\label{whataregime}
  J=O\left(\sqrt{\lambda }\right),\qquad k=O(1).
\end{equation}
The two-point function $\left\langle \mathcal{O}^\dagger _{J}\mathcal{O}_J\right\rangle$ then is describes by the classical string solution \cite{Tsuji:2006zn,Janik:2010gc}:
\begin{eqnarray}\label{AdSgeod}
 x&=&R\tanh\kappa \tau ,\qquad R=\frac{|x_1-x_2|}{2}
\nonumber \\
 z&=&\frac{R}{\cosh\kappa \tau }
 \\  \label{Sgeod}
\varphi  &=&i\kappa \tau ,\qquad \theta =\frac{\pi }{2}\,.
\end{eqnarray}
which is basically the Euclidean continuation of the BMN geodesic \cite{Berenstein:2002jq}. The parameter $\kappa $ is related to the R-charge/dimension of the fat operators:
\begin{equation}\label{}
 \kappa =\frac{J}{\sqrt{\lambda }}\,.
\end{equation}

Substituting this solution into (\ref{opeofcpo}) we find  that the
OPE coefficient scales with the distance in the right way:
$\mathbb{C}_k[\mathcal{O}^\dagger
_{J+k}(x_1)\mathcal{O}_J(x_2)]\propto R^k=2^{-k}|x_1-x_2|^k$, which
agrees with the conformal structure of the three-point function.
Pulling out the factor of $|x_1-x_2|^k$  according to
(\ref{andope}), and performing integration over the classical
worldsheet we find:
\begin{equation}\label{iral}
 C^{J+k}_{J,k}=\frac{1}{N}\,2^{-{k}-1}J(k+1) \sqrt{k}\,\kappa\int_{-\infty }^{+\infty }d\tau \,\,
 \frac{\, {\rm e}\,^{-k\kappa \tau }}{\cosh^{k+2}\kappa \tau }
 =\frac{1}{N}\,J\sqrt{k},
\end{equation}
in agreement with the exact result -- in the regime (\ref{whataregime}) we cannot distinguish $J$ and $J+k$.

\subsection{Spinning strings in $S^5$}

When the fat operators are dual to a spinning string on $S^5$, the AdS part of the solution for the two-point function is again described by the point-like geodesic (\ref{AdSgeod}). In $S^5$ the worldsheet is a periodic solution of the $O(6)$ sigma-model $\mathbf{n}(\sigma ,\tau )$.
The OPE coefficient with the BMN  chiral primary is given
by (\ref{opeofcpo}), with (\ref{hw}) substituted for the spherical function:
\begin{equation}\label{kcpo}
 \mathbb{C}_k=
 \frac{(k+1)\sqrt{k\lambda }}{8\pi N}\int_{}^{}d^2\sigma \, z^k\,{\rm e}\,^{ik\varphi }\sin^k\theta
 \left[
 \frac{\left(\partial x\right)^2-\left(\partial z\right)^2}{z^2}-
 \left(\partial \mathbf{n} \right)^2\right].
\end{equation}
Using the explicit form of the AdS geodesic and excluding $(\partial \mathbf{n}/\partial \sigma )^2$ with the help of the Virasoro constraints, we find, after extracting the factor of $|x_1-x_2|^k$:
\begin{equation}\label{C_k}
 C_k=-\frac{\left(k+1\right)\sqrt{k\lambda }}{2^{k+2}\pi N}
 \int_{-\infty }^{+\infty }\frac{d\tau }{\cosh^k\kappa \tau }\,
 \int_{0}^{2\pi }d\sigma \,\, {\rm e}\,^{ik\varphi }\sin^k\theta \left[
 \kappa ^2\tanh^2\kappa \tau +\left(\frac{\partial \mathbf{n}}{\partial \tau }\right)^2
 \right].
\end{equation}

The periodic solutions of the $O(6)$ sigma-model  can be constructed
in full generality with the help of integrability methods
\cite{Kazakov:2004qf,Beisert:2004ag}. The dual operators can then be
identified through the Bethe ansatz equations. In principle the
general finite-gap solution is know in a relatively explicit form
\cite{Dorey:2006zj}, but not explicit enough to calculate the
integral (\ref{C_k}). Here we will compute the OPE coefficient for
the folded string solution. The starting point is the string that
rotates on the big circle of $S^5$ and spins around its centre of
mass \cite{Frolov:2003xy}, which then should be Wick rotated to the
Euclidean signature. The dual operator corresponds to the two-cut
solution of the classical Bethe equations \cite{Beisert:2003xu}.

The folded string solution is characterized by  two frequencies
$\omega _1$, $\omega _2$ of rotation around $S^5$  and around the
string's centre of mass:
\begin{equation}\label{folded}
 \varphi  =i\omega_1 \tau ,\qquad \psi =i\omega _2\tau ,\qquad \alpha
 =\frac{\pi }{2}\,,\qquad \theta =\theta (\sigma ),\qquad
 \acute{\theta }^2=\kappa ^2-\omega _1^2\sin^2\theta -\omega
 _2^2\cos^2\theta.
\end{equation}
The $\sigma $ dependence can be integrated in terms of elliptic functions whose
modulus $s$ is related to the frequencies  by
\begin{equation}\label{sdef}
 s=\frac{\kappa ^2-\omega _1^2}{\omega _2^2-\omega _1^2}\,.
 \end{equation}
The periodicity in  $\sigma $ relates $\omega _1$, $\omega _2$ and $\kappa $:
 \begin{equation}\label{virs}
 \sqrt{\omega
 _2^2-\omega _1^2}=\frac{2K(s)}{\pi }\,,
\end{equation}
where $K(s)$ is the complete elliptic integral of the first kind.

Instead of the frequencies, it is more convenient to characterize
the solution by the conserved quantum numbers, the two angular
momenta and the energy:
\begin{equation}\label{}
 J_1=\frac{\sqrt{\lambda }\omega _1E(s)}{K(s)}\,,\qquad
 J_2=\frac{\sqrt{\lambda }\omega _2\left(K(s)-E(s)\right)}{K(s)}\,,\qquad
 \Delta =\sqrt{\lambda }\kappa ,
\end{equation}
where $E(s)$ is the complete
elliptic integral of the second kind.
The angular momenta $J_1$ and $J_2$ are the Noether charges associated with the
isometries $\varphi \rightarrow \varphi +\,{\rm const}\,$, $\psi
\rightarrow \psi +\,{\rm const}\,$. The dual operator has the form $\mathop{\mathrm{tr}}Z^{J_1}W^{J_2}+{\rm permutations}$,  $Z=\Phi _1+i\Phi _2$, $W=\Phi _3+i\Phi _4$, and $\Delta $ is its exact scaling dimension. Eqs.~(\ref{sdef}) and (\ref{virs}) express the dimension as a function of the R-charges in an implicit form: $\Delta =\Delta (J_1,J_2)$. Finally, it is useful to introduce a parameter
\begin{equation}\label{}
 a=\frac{\omega _1}{\kappa }=\frac{J_1K(s)}{\Delta E(s)}\,.
\end{equation}

The the integral in (\ref{C_k}) can be calculated in terms of the hypergeometric function:
\begin{eqnarray}\label{cfold}
 C^{\rm fold.}_{{\rm fold.},k}
 &=&\frac{1}{N}\,\,
 \frac{\pi \sqrt{k}\Delta \left(1-a^2\right)\Gamma \left(\frac{\left(1+a\right)k}{2}\right)
 \Gamma \left(\frac{\left(1-a\right)k}{2}\right)}{8sK(s)(k-1)!}
 \nonumber \\ &&\times
 \left[
 (k+1-s){}_2F_1\left(-\frac{k-1}{2}\,,\frac{1}{2}\,;1;s\right)
 -(k+1){}_2F_1\left(-\frac{k+1}{2}\,,\frac{1}{2}\,;1;s\right)
 \right].
\end{eqnarray}
At $a\rightarrow 1$, the folded string shrinks to a point and goes
over to the point-like solution dual to the chiral primary operator.
However,  we do not recover the OPE coefficient of three chiral
primaries (\ref{iral}) in the limit $a\rightarrow 1$. There is an
extra factor of $(k-1)/2k$. Why? The difference can be attributed to
the anomaly that arises when the slim string vertex operator
approaches the fat string vertex operator at $\tau = -\infty $. The
integrand in (\ref{kcpo}) contains an exponential factor $\,{\rm
e}\,^{ik\varphi }z^k\sim \,{\rm e}\,^{k(\kappa -\omega _1)\tau }$
that cuts off the integral at large negative $\tau $, because
$\kappa >\omega _1$. In the point-string limit $\kappa \rightarrow
\omega _1$ the suppression disappears. The square bracket decomposes
in the limit in two terms. One is additionally suppressed as $\,{\rm
e}\,^{2\kappa \tau }$ and survives the BMN limit, making the $\tau $
integral in (\ref{iral}) manifestly convergent. The other term is
proportional to $\acute{\theta }^2\sim \kappa -\omega _1$ and
vanishes as the string becomes point-like, but it does not have the
additional exponential suppression and integration in $\tau $
diverges in the limit leading to the $0/0$ cancelation.

\subsection{Saddle point approximation}

At large $k$ the integral over the vertex operator insertion in
(\ref{kcpo}) is dominated by a saddle point, a solution of the
equation
\begin{equation}\label{sstar}
 \cot\theta \,\partial _{\mathbf{a}}\theta  +i\partial _{\mathbf{a}} \varphi +\frac{\partial _{\mathbf{a}}
 Z}{Z}=0.
\end{equation}
This constitutes a set of two equations on two variables and consequently is satisfied in a finite number of points.
The OPE coefficient then is exponentially suppressed:
\begin{equation}\label{}
 C_k\sim \,{\rm e}\,^{-kS},
\end{equation}
where $S=-\ln\sin\theta (\sigma _*)-i\varphi (\sigma _*)-\ln
Z(\sigma _*)$ and $\sigma _*=(\tau _*,\sigma _*)$ is the solution
of (\ref{sstar}) with the smallest possible suppression.

For instance, the folded string OPE with the chiral primary is
exponentially small
 at large $k$:
\begin{equation}\label{}
 C^{\rm fold.}_{{\rm fold.},k}
 \simeq\frac{1}{N}\,\left(\frac{1}{s}-1\right)\sqrt{\frac{\lambda }{k}}
 \exp\left[k\left(
 \frac{1+a}{2}\,\ln\frac{1+a}{2}+\frac{1-a}{2}\,\ln\frac{1-a}{2}
 \right)\right]
\end{equation}
For the folded string there
are two degenerate  saddle points:
\begin{equation}\label{sadfol}
 \tau _*=\frac{1}{2\kappa }\,\ln\frac{1-a}{1+a}\,,\qquad
 \sigma _*=\frac{\pi }{2}\,{\rm ~and~}\frac{3\pi }{2}\,:~\theta (\sigma
 _*)=\frac{\pi }{2}\,.
\end{equation}
The pre-exponential factor vanished at the saddle point
and has to be expanded to the second order to get the correct
coefficient in front. It is actually easier to compute the integral exactly and then take the large-$k$ limit. In the point-like string limit of
$a\rightarrow 1$, the saddle point (\ref{sadfol}) approaches the
boundary and the exponential suppression disappears.

\begin{figure}[t]
\centerline{\includegraphics[scale=0.4]{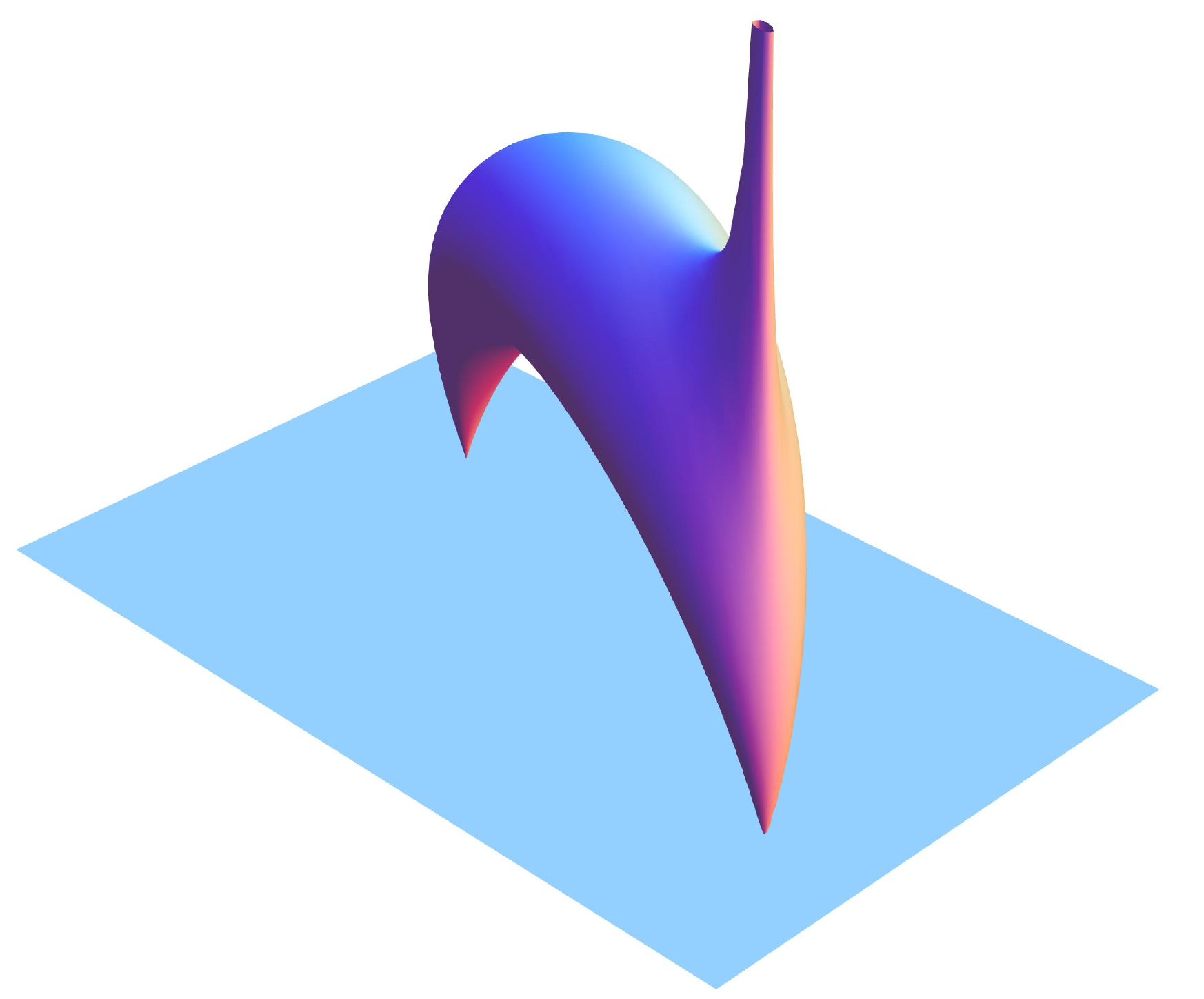}}
\caption{\label{figspike}\small The vertex operator creates an
infinite spike.}
\end{figure}

Our calculations were accurate at $k=O(1)$, but the limit of
$k\rightarrow \infty $ should have an overlapping region of validity
$\sqrt{\lambda }\gg k\gg 1 $ with the $k/\sqrt{\lambda }\rightarrow
0$ limit of the OPE with all three string states being
semiclassical\footnote{This has been explicitly verified for the
Wilson loop correlator with a local operator
\cite{Zarembo:2002ph,Pestun:2002mr}.}. Because we took the limit $x_3\rightarrow \infty $ in the three-point function, the vertex operator of a fat string with $k\sim\sqrt{\lambda }$
creates
an infinite spike on the
worldsheet with two other vertex insertions ending on the boundary of AdS, fig.~\ref{figspike}.
When
$k/\sqrt{\lambda }\rightarrow 0$ the spike shrinks to zero size. The
saddle point $\sigma _*$ determines the position where the spike is
attached to the macroscopic worldsheet, or better to say, the position where
the equations of motion allow the spike to appear
when we gradually raise $k/\sqrt{\lambda }$ from zero. We are going to demonstrate this explicitly in the next section, by starting with finite $k/\sqrt{\lambda }$ and then taking the limit of $k/\sqrt{\lambda }\rightarrow 0$.

\section{Fine structure of the spike}

Now we consider the regime when we insert the vertex operator
(\ref{hw}) with $k\sim \sqrt{\lambda }$  in the string path
integral. When $k\sim\sqrt{\lambda }$ the vertex operators changes
the boundary conditions for the embedding coordinates and creates an
infinite spike \cite{Zarembo:2002ph,Pestun:2002mr}
(fig.~\ref{figspike}).
 There may be other vertex operators or Wilson loops, which we do not specify, as we will be interested
in the structure of the solution in the vicinity of the insertion
point. The local worldsheet coordinates will be denoted by $w$,
$\bar{w}$, and we will assume that the vertex operator is inserted
at $w=0$.

At $\lambda \rightarrow \infty $, the logarithm of the vertex
operator is of the same order as the action\footnote{The action is
written in the conformal gauge. The dependence on the 2d metric
should also drop from the vertex operator due to its marginality.}:
\begin{equation}\label{}
 S=\frac{\sqrt{\lambda }}{4\pi }\int_{}^{}d^2w
 \left[
 \frac{\left(\partial z\right)^2+\left(\partial x\right)^2}{z^2}+
 \left(\partial\theta  \right)^2+\sin^2\theta \,\left(\partial \varphi
 \right)^2
 \right]
 -k\ln z(0)-ik\varphi (0)-k\ln\sin\theta (0).
\end{equation}
We do not display other coordinates or other vertex operators as we will be interested in the most singular behavior near $w=0$.

The vertex operator creates a localized source in the equations of
motion:
\begin{eqnarray}\label{eqm}
 \partial_{\mathbf{a}} \left(\frac{\partial ^{\mathbf{a}}x^\mu}{z^2}\right) &=&0\nonumber \\
 -\partial ^2\ln z-\frac{\left(\partial x\right)^2}{z^2}&=&2\pi \chi \delta
 (w)\nonumber \\
 -\partial ^2\theta +\sin\theta \,\cos\theta \,\left(\partial \varphi
 \right)^2&=&2\pi \chi \cot\theta \,\delta (w)\nonumber \\
 -\partial _{\mathbf{a}}\left(\sin^2\theta \,\partial ^{\mathbf{a}}\varphi \right)&=&2\pi i\chi \delta
 (w),
\end{eqnarray}
where
\begin{equation}\label{}
 \chi =\frac{k}{\sqrt{\lambda }}\,.
\end{equation}
The source determines the boundary conditions for the worldsheet
fields at $w=0$:
\begin{eqnarray}\label{b.c.}
 \ln z&\rightarrow& -\chi \ln|w|\nonumber \\
 \theta &\rightarrow &\frac{\pi }{2}\nonumber \\
 \varphi &\rightarrow &-i\chi \ln|w|.
\end{eqnarray}
Let us see in more detail what is the structure of the classical
worldsheet in the vicinity of $w=0$.

\begin{figure}[t]
\centerline{\includegraphics[scale=0.4]{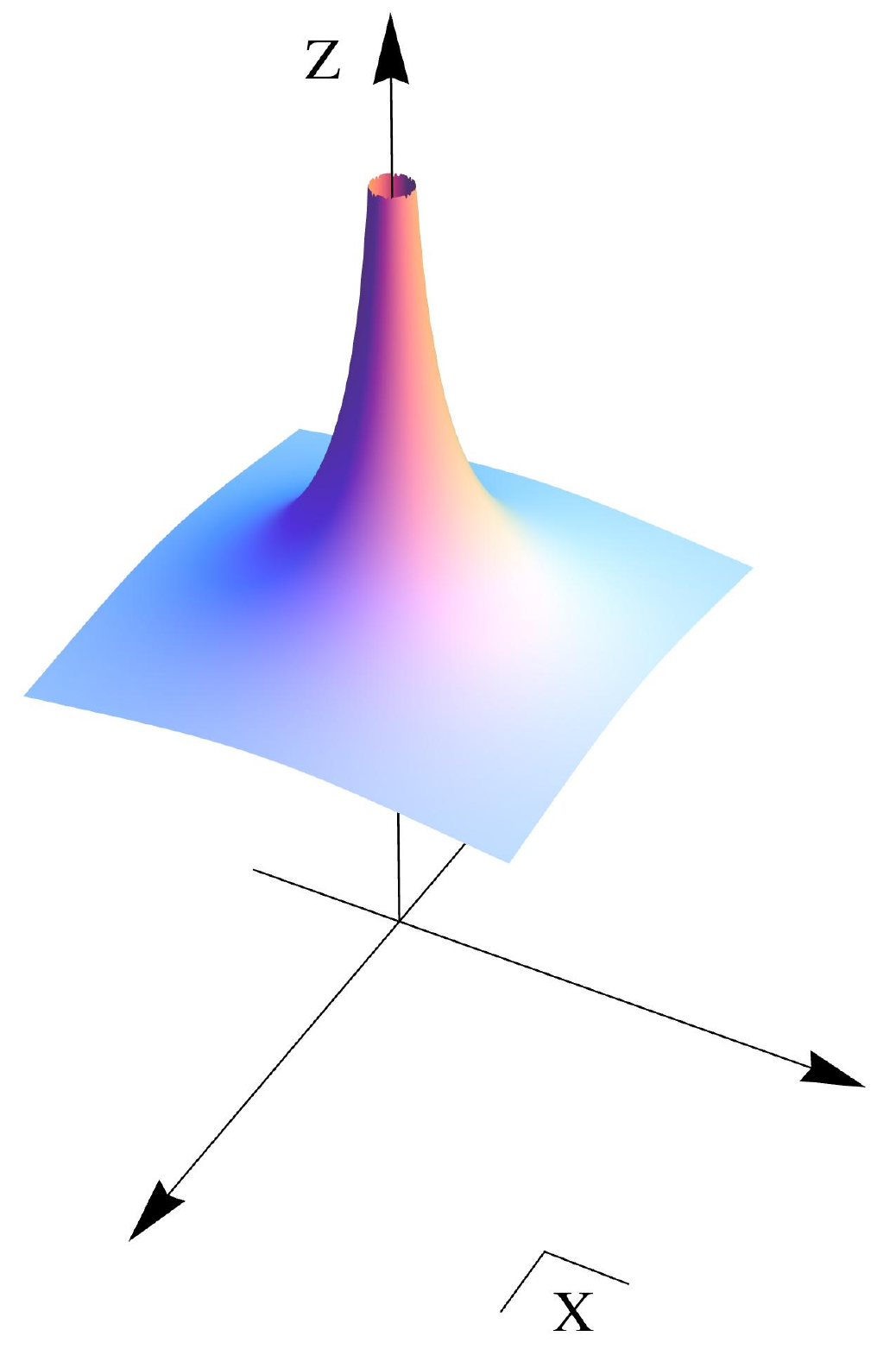}}
\caption{\label{finespike}\small The string worldsheet in the
vicinity of a spike.}
\end{figure}

We begin with the shape of the string in AdS. According to the
boundary conditions, the worldsheet has an infinite spike going all
the way to the horizon (fig.~\ref{finespike}),
with the AdS $z$ coordinate scaling as $z\sim |w|^{-\chi }$. As far as the $x^\mu $ coordinates are concerned, we can try
an axially symmetric scaling ansatz $x\sim |w|^\alpha w$, where $x=x^1+ix^2$ is the complex coordinate on the boundary of $AdS_3$.
 The ansatz goes through
the equations if $\alpha $ satisfies:
$$
 \frac{\alpha }{2}\left(\frac{\alpha }{2}+1+\chi \right)+\left(\frac{\alpha }{2}+1\right)
 \left(\frac{\alpha }{2}+\chi \right)=0,
$$
giving
\begin{equation}\label{}
 \alpha =\sqrt{1+\chi ^2}-1-\chi .
\end{equation}

It is easy to understand the meaning of this result by making the conformal map to the cylinder:
\begin{equation}\label{polar}
 w=\,{\rm e}\,^{-\tau +i\sigma }.
\end{equation}
The transformation $z\rightarrow \,{\rm e}\,^{\chi \tau }z$,
$x\rightarrow \,{\rm e}\,^{-\chi \tau }x$ generates a mass term of
magnitude $\chi $ in the equations of motion for $x$. The frequency
of the $n$th Fourier mode of a 2d field of mass $\chi $ is equal to
$\omega _n=\sqrt{n^2+\chi ^2}$. The axially symmetric solution
corresponds to the single-winding mode
 $x\sim \,{\rm e}\,^{-\omega _1\tau
+i\sigma }$.

The rate at which $x$ goes to zero at $w\rightarrow 0$ is always larger than the rate at which $\ln z$ goes to infinity. For this reason $x$  does not
backreact on $z$ up to the next-to-next-to-leading order.  It is therefore
easy to deduce the next term in the expansion of $z$ in $w$ by just
treating $\ln z$ as a free massless field with the source at the origin. The most general solution  then reads:
\begin{eqnarray}\label{spikeads}
 z&=&|w|^{-\chi }\left(z_0+\partial z_0w+\bar{\partial
 }z_0\bar{w}\right)+\ldots \nonumber \\
 x^\mu &=&x^\mu _0+|w|^{\sqrt{1+\chi ^2}-1-\chi }
 \left(\partial x^\mu _0 w+\bar{\partial }x^\mu _0\bar{w}\right)+\ldots
 ,
\end{eqnarray}
where $z_0$, $x^\mu _0$, $\partial z_0$, $\bar{\partial }z_0$,
$\partial x^\mu _0$, and $\bar{\partial }x^\mu _0$ are
constants,  not determined by the equations of motion.

Now we can take the $\chi \rightarrow 0$ limit. The spike then
disappears, and (\ref{spikeads}) becomes an ordinary Taylor
expansion of the regular solution without the source at $w=0$. We
can thus regard $ (z_0,x^\mu _0)$ as the target-space coordinates of
the point at which the spike is attached to the string worldsheet.
The constants $\partial z_0 $, ...  are Taylor expansion
coefficients of the smooth solution $(z_0(w),x^\mu _0(w))$ for the
string without the spike.

We now turn to the $S^5$. The boundary conditions imply that $\,{\rm
e}\,^{i\varphi }\rightarrow \,{\rm const}\,|w|^{\chi }$ and
$\cos\theta \rightarrow \,{\rm const}\,|w|^{\chi }$ at $w\rightarrow
0$. Corrections to this behavior are of two types. There are regular
corrections in powers of $w$, $\bar{w}$, as well as non-analytic
corrections in $|w|^\chi $. Which of those are more important
depends on the value of $\chi $. When $\chi \rightarrow 0$ and the
spike shrinks leaving behind some regular solution $\theta _0(w)$,
$\varphi_0 (w)$, the non-analytic corrections are more important,
moreover in the strict $\chi \rightarrow 0$ limit they cease to be
suppressed and have to be re-summed. It is possible to do the
re-summation explicitly, although the details are somewhat lengthy.
The solution which is accurate to all orders in $|w|^\chi $ and to
the first two orders in $w$, $\bar{w}$ is derived in the
appendix~\ref{appendix}:
\begin{eqnarray}\label{spikes}
 \,{\rm e}\,^{i\varphi }&=&\frac{|w|^\chi \,{\rm e}\,^{i\varphi _0}\sin\theta _0}
 {\sqrt{1-|w|^{2\chi }\cos^2\theta _0}}\left[
 1+i\,\frac{1-|w|^\chi \cos^2\theta _0}{1-|w|^{2\chi }\cos^2\theta _0}
 \left(\partial \varphi _0w+\bar{\partial }\varphi _0\bar{w}\right)
 \right. \nonumber \\ &&\left.
 +\frac{\left(1-|w|^\chi \right)\cot\theta _0}{1-|w|^{2\chi }\cos^2\theta _0}
 \left(\partial \theta _0w+\bar{\partial }\theta _0\bar{w}\right)
 \right]+\ldots\nonumber \\
 \cos\theta &=&|w|^\chi \cos\theta _0-|w|^{\sqrt{1+\chi ^2}-1}
 \left[
 \frac{1-|w|^\chi \cos^2\theta _0}{\sin\theta _0}\left(\partial \theta _0w
 +\bar{\partial }\theta _0\bar{w}\right)
 \right. \nonumber \\ &&\left.
 +i\left(1-|w|^\chi \right)\cos\theta _0\left(\partial \varphi _0w
 +\bar{\partial }\varphi _0\bar{w}\right)
 \right]+\ldots ,
\end{eqnarray}
where again $\theta _0$, $\varphi _0$, $\partial \theta _0$, ... are
 constants not fixed by the equations of motion. These constants
can be regarded as the Taylor coefficients of the regular solution.
The above approximate solution is valid in the limit of
$w\rightarrow 0$ (in that case all subleading $|w|^{n\chi }$ terms
can be dropped), as well as in the double-scaling limit
$w\rightarrow 0$, $\chi \rightarrow 0$, with $|w|^\chi $ fixed (in
which case $|w|^{\sqrt{1+\chi ^2}-1}$ should  be replaced by $1$).

To the leading order in $w$, the solution (\ref{spikeads}),
(\ref{spikes}) automatically satisfies the Virasoro constraints,
which reflects the marginality of the vertex operator (\ref{hw}). At
the next order the Virasoro constraints are not automatic and lead
to the relationship between the expansion coefficients $\partial
z_0$, ...

Requiring that
\begin{equation}\label{}
 0=T_{ww}\equiv
 \left( \frac{\partial z}{z}\right)^2+\left(\partial \theta
 \right)^2+\sin^2\theta \,\left(\partial \varphi \right)^2,
\end{equation}
and imposing the same condition on $T_{\bar{w}\bar{w}}$, we find a
linear relationship between the coefficients:
\begin{equation}\label{}
 \frac{\partial z_0}{z_0}+\cot\theta _0\,\partial \theta _0+i\partial \varphi
 _0=0,\qquad
 \frac{\bar{\partial} z_0}{z_0}+\cot\theta _0\,\bar{\partial} \theta _0+i\bar{\partial}
  \varphi
 _0=0.
\end{equation}
Viewed as a condition on the unperturbed solution $\theta _0$,
$\varphi _0$, it determines a point on the worldsheet to which the
spike can be attached without violating the Virasoro constraints.
This condition is the same as the saddle point equation
(\ref{sstar}), which was derived when we neglected the backreaction of the vertex operator on the
shape of the string.

\section{Discussion}

It would be interesting to compare  the three-point functions
computed at weak and at strong coupling, in particular to check if
the structural observations made recently in the one-loop
corrections to the three-point functions of scalar operators
\cite{Grossardt:2010xq} survive at strong coupling. It would also be
interesting to generalize integrability methods to three-point and
higher correlation functions.  The spectral equations derived using
integrability methods determine the eigenvalues of the light-cone
string Hamiltonian. To compute correlation functions, which are
analogous to closed string amplitudes in the familiar flat-space
setting, one also needs to know the wavefunctions, which are more or
less equivalent to vertex operators. The vertex operators of the
chiral states can be deduced from the supergravity equations of
motion, and probably do not receive $\alpha '$ ($1/\sqrt{\lambda }$)
corrections. It would be interesting to understand how to
systematically construct vertex operators of non-protected states
\cite{Tseytlin:2003ac,Buchbinder:2010vw}. Then one will be able to
compute their holographic three-point functions, at least in the
approximation used in this paper.

\subsection*{Acknowledgments}
I would like to thank R.~Janik and V.~Kazakov for collaboration on
the initial stages of this project and for numerous discussions. I
benefited a lot from the discussions with J.~Escobedo, I.~Kostov,
D.~Serban, A.~Sever and P.~Vieira.  I would like to thank M.~Costa,
R.~Monteiro, J.~Santos and D.~Zoakos for sending me the draft of
\cite{Costa:2010rz} prior to publication, and the Perimeter
Institute for hospitality during the course of this project. This
work was supported in part by the Swedish Research Council under
grant 621-2007-4177, in part by the ANF-a grant 09-02-91005, and in
part by the grant for support of scientific schools NSH-3036.2008.2.

\appendix

\section{Spike in $S^5$}\label{appendix}

It is easier to analyze the solution in the polar coordinates
(\ref{polar}). To the leading order, the solution depends only on
$\tau  $, and the equations of motion become ordinary differential equations:
\begin{eqnarray}\label{}
 \left(\sin^2\theta \,\dot{\varphi } \right)\dot{}&=&0
 \nonumber \\
 \ddot{\theta }-\sin\theta \,\cos\theta \,\dot{\varphi }^2&=&0.
\end{eqnarray}
The boundary conditions (\ref{b.c.}) require that $\varphi
\rightarrow i\chi \tau $, $\theta \rightarrow \pi /2$ at $\tau
\rightarrow \infty $. Taking this into account we find the solution
that depends on the two integration constants:
\begin{eqnarray}\label{linear1}
 \cos\theta &=&\,{\rm e}\,^{-\chi \tau }\cos\theta _0
 \\
 \label{linear2}\,{\rm e}\,^{i\varphi }&=&\,{\rm e}\,^{i\varphi _0}\cot\theta .
\end{eqnarray}

The next order  in $w$, $\bar{w}$ corresponds to the first
Kaluza-Klein mode in $\sigma $ and is suppressed by an extra factor
of $\,{\rm e}\,^{-\tau }$, which is a small correction at $\tau
\rightarrow \infty $. The equations of motion can be consequently
linearized.

We consider a slightly more general setup when the linear correction
is the $n$th harmonic. Substituting
$$\delta \theta =\xi \,{\rm
e}\,^{-n\tau +in\sigma }, \qquad \delta\varphi =\eta \,{\rm
e}\,^{-n\tau +in\sigma }$$
into the equations of motion (\ref{eqm})
and linearizing in $\xi $ and $\eta$, we get:
\begin{eqnarray}\label{}
 -\left(\sin^2\theta \,\dot{\eta }\right)\dot{}+n\sin^2\theta \,\dot{\eta }
 +n\left(\sin^2\theta \,\eta \right)\dot{}
 +2i\chi n\cot\theta \,\xi -2i\chi \left(\cot\theta \,\xi \right)\dot{}&=&0
 \nonumber \\
 -\ddot{\xi }+2n\dot{\xi }+\chi ^2\left(1-\cot^4\theta \right)\xi +2i\chi \cot\theta
 \,\dot{\eta }-2i\chi n\cot\theta \,\xi &=&0,
\end{eqnarray}
where $\cos\theta $ is the leading-order solution (\ref{linear1}).

There are two possible
regimes: (I) $\tau \rightarrow \infty $, finite $\chi $ and (II) $\tau \rightarrow
\infty $, $\chi \rightarrow 0$, $\chi \tau $-fixed. In case (I)
we can set $\theta =\pi /2$, which yields
\begin{equation}\label{}
 \xi =\,{\rm const}\,\,{\rm e}\,^{\left(n-\sqrt{n^2+\chi ^2}\right)\tau
 },
 \qquad
 \eta =\,{\rm const}\,.
\end{equation}

In case (II) there is a scaling region $\tau \sim
1/\chi $, where $\xi =\xi (\chi \tau )$ and $\eta =\eta (\chi \tau )$.
We can then  neglect the terms of order $\chi ^2$ (and keep those of
order $\chi $) in the linearized equations. In particular we can
neglect the second derivatives in $\tau $. The equations then are of the first order and can be
integrated:
\begin{eqnarray}\label{regII}
 \xi &=&\sin\theta \left(\frac{C_1}{1-\cos\theta }+\frac{C_2}{1+\cos\theta }\right)
 \nonumber \\
 \eta &=& i \left(\frac{C_1}{1-\cos\theta }-\frac{C_2}{1+\cos\theta }\right)
\end{eqnarray}
The constants of integration $C_1$, $C_2$ can be expressed in terms
of  $\partial \theta _0$, $\partial \varphi _0$ by setting $\chi
=0$, when $\theta \rightarrow \theta _0$ and $\xi \rightarrow
\partial \theta _0$, $\eta\rightarrow \partial \varphi _0$. Finally,
we can write the solution of the linearized equations in  the form
that is valid both in the regime (I) and (II) by simply multiplying
$\xi $ in (\ref{regII}) by $\,{\rm e}\,^{\left(n-\sqrt{n^2+\chi
^2}\right)\tau
 }$, which is equivalent to $1$ in the regime (II). This gives the solution
 (\ref{spikes}) quoted in the main text.

\bibliographystyle{nb}
\bibliography{refs}

\end{document}